\documentclass[aps,prl,showpacs,twocolumn]{revtex4}
\usepackage{graphicx}
\usepackage{bm}
\usepackage{amsmath}
\usepackage{amsfonts}

\def\ra{\rangle}
\def\la{\langle}
\def\be{\begin{equation}}
\def\ee{\end{equation}}
\def\bea{\begin{eqnarray}}
\def\eea{\end{eqnarray}}

\def\pa{\partial}

\def\dag{^\dagger}

\def\al{\alpha}

\begin{document}
\title{Theory of the Thermal Hall Effect in Quantum Magnets}

\author{Hosho Katsura$^1$, Naoto Nagaosa$^{1,2}$, Patrick A Lee$^3$}
\affiliation{%
$^1$Cross Correlated Materials Research Group, Frontier Research System, 
Riken,2-1 Hirosawa, Wako, Saitama 351-0198, Japan\\
$^2$Department of Applied Physics, The University of Tokyo,
7-3-1, Hongo, Bunkyo-ku, Tokyo 113-8656, Japan\\
$^3$Department of Physics, Massachusetts Institute of Technology,
77 Massachusetts Avenue, Cambridge, MA 02139}
\begin{abstract}
We present a theory of the thermal Hall effect in insulating quantum magnets,
where the heat current is totally carried by charge-neutral objects 
such as magnons and spinons. 
Two distinct types of thermal Hall responses are identified. 
For ordered magnets, the intrinsic thermal Hall effect for magnons 
arises when certain conditions are satisfied for the lattice geometry and the 
underlying magnetic order. 
The other type is allowed in a spin liquid which is a novel quantum state since there is no order even at zero temperature.  
For this case, the deconfined spinons contribute 
to the thermal Hall response due to Lorentz force. These results offer a clear 
experimental method to prove the 
existence of 
the deconfined spinons via a thermal transport phenomenon.
\end{abstract}
\pacs{71.10.Hf, 72.20.-i, 75.47.-m}

\maketitle
The ground state and low energy excitations of correlated electronic systems
are the subject of recent intensive interests, and especially the possible 
quantum liquid states are the focus both theoretically and 
experimentally~\cite{Shimizu_PRL, S_Yamashita, Okamoto_PRL_hyperkagome, Helton_PRL, Motrunich2}. 
For the quantum magnets, 
magnetic susceptibility, neutron scattering 
and specific heat are the experimental tools to study this issue. 
In the conducting systems, on the other hand, 
charge transport properties also offer important clues to the novel  electronic states such as the non-Fermi liquid or the quantum Hall liquid. 
Therefore, a natural question is whether 
there are any transport properties 
in insulating quantum magnets 
which provide insight into the ground state.
To answer this question, we study in this Letter the thermal Hall effect theoretically and find several different mechanisms leading to the classification of the quantum magnets.    


For 
a finite Hall response, 
time-reversal
symmetry must be broken due to 
magnetic field and/or 
magnetic ordering. 
The Hall effect in 
itinerant magnets, 
where the spin structure and conduction electron motion are coupled,
has been studied extensively. In this case, in addition to the
usual Lorentz force, the scalar (spin) chirality defined for three spins as 
$\vec S_i \cdot (\vec S_j \times \vec S_k)$ plays an important 
role~\cite{Taguchi_science,Ohgushi-Murakami-Nagaosa, Nagaosa_JPSJ}. 
The scalar chirality acts as a fictitious magnetic flux for the conduction 
electrons and gives rise to a non-trivial topology of 
the Bloch wave functions, leading to the Hall effect. 
It is natural to expect that a similar effect occurs even in the 
localized spin systems for e.g. the spin current~\cite{Fujimoto}.
Another important tool to detect charge-neutral modes is 
the thermal transport measurement. 
In low-dimensional magnets, the ballistic thermal transport 
property was predicted from the integrability 
of the one-dimensional Heisenberg model \cite{Zotos_PRB} and 
has been experimentally observed in Sr$_2$CuO$_3$ \cite{Sologubenko_PRB}. 
In $\kappa$-(ET)$_2$Cu$_2$(CN)$_3$, one of possible 
candidates for two-dimensional quantum spin 
liquids~\cite{Shimizu_PRL}, the thermal transport measurement was used as a probe 
to unveil the nature of low-energy spin 
excitations \cite{M_Yamashita}. 
The measurements have been limited to the 
longitudinal thermal conductivity so far.
In this paper we predict a  
non-zero thermal Hall conductivity, i.e., the Righi-Leduc effect
which will provide 
important information as described below.
 

First, we need to consider the influence of the external magnetic
field on localized spin systems.
In addition to the Zeeman coupling, we have the ring exchange process
leading to the coupling between the scalar chirality and 
external magnetic fields.
This coupling is derived from the $t/U$ expansion for the Hubbard model at 
half filling with on-site Coulomb interaction $U$ and complex hopping 
$t_{ij}=t e^{iA_{ij}}$ \cite{Sen_Chitra, Motrunich1} and its explicit 
form is given by
\begin{equation}
H_{\rm ring} = -\frac{24t^3}{U^2} \sin \Phi~ \vec S_i \cdot (\vec S_j \times \vec S_k),
\label{ring_exchange}
\end{equation}
where $\Phi$ is the magnetic flux through the triangle formed by the 
sites $i$, $j$, and $k$ in a counterclockwise way. Since the 
coefficient is proportional to $t^3/U^2$, it is expected to 
be small. In the vicinity of the Mott 
transition, however, this coupling is not negligible. 
We first examine the effect of $H_{\rm ring}$ 
within the spin-wave approximation. 
Then we find that if the lattice geometry and the magnetic order satisfy certain conditions, 
the magnons can experience the fictitious magnetic field and there occurs the intrinsic thermal 
Hall effect, i.e., the thermal Hall conductivity $\kappa^{xy}$ due to the anomalous velocity of the magnons.
In this case, $\kappa^{xy}$ is independent of the lifetime of magnons ($\tau$), whereas the longitudinal one $\kappa^{xx}$ depends on $\tau$~\cite{Sinitsyn_JPCond, extrinsic}. 
It can be regarded as a bosonic analogue of the quantum Hall effect with 
zero net flux~\cite{Haldane_PRL}. 
We also derive a TKNN-type formula \cite{TKNN} of the thermal Hall 
conductivity for a general free-bosonic Hamiltonian. 
It 
should 
be possible to apply this formula to the recently found 
phonon thermal Hall effect \cite{Rikken_PRL}.  
Finally, we consider the effect of $H_{\rm ring}$ in quantum spin 
liquids. Since there is no magnetic order in such a system, 
it 
has been proposed 
that 
deconfined fermionic spinons 
exist. 
In contrast to the magnons, the spinons, which are the
gauge dependent object, can feel the 
vector potential $\vec A$ just as in the case of electrons, 
leading to the Landau level formation~\cite{Motrunich1}. 
We propose a novel way to detect the spinon deconfinement 
via the thermal Hall effect measurement in a 
candidate of quantum spin liquid, $\kappa$-(ET)$_2$Cu$_2$(CN)$_3$.


{\it NO-GO theorem for the coupling to magnetic flux}. ---
Let us first consider the spin-wave expansion of Eq. (\ref{ring_exchange})
to find a system in which the intrinsic thermal Hall effect occurs.  
We consider the collinear ground state spin configurations. 
The fluctuation of the scalar chirality up to the second order in 
$\delta {\vec S} \equiv \vec S-\langle \vec S \rangle$ is written as
\begin{equation}
\langle \vec S_i \rangle \cdot (\delta \vec S_j \times \delta \vec S_k) +
\langle \vec S_j \rangle \cdot (\delta \vec S_k \times \delta \vec S_i) +
\langle \vec S_k \rangle \cdot (\delta \vec S_i \times \delta \vec S_j),
\label{chirality}
\end{equation}
where $\langle S^a \rangle$ denotes the $a$-component ($a=x, y$, or $z$) 
of the ordered moment. 
Note here that the linear order terms in $\delta {\vec S}$ vanish 
since $\langle {\vec S}_i \rangle \times \langle {\vec S}_j \rangle = {\vec 0}$.
\begin{figure}[tb]
\begin{center}
\vspace{.5cm}
\hspace{-.0cm}\includegraphics[width=0.9\columnwidth]{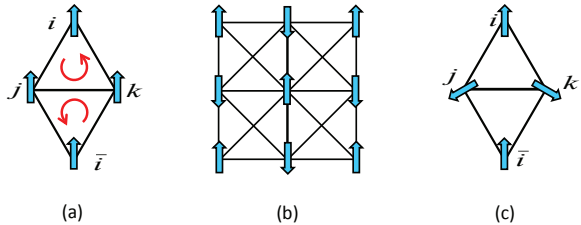}
\vspace{.0cm}
\caption{(a) A portion of the triangular lattice with ferromagnetic order. 
The blue arrows indicate the directions of spins. The counterclockwise rotations (red arrows) 
indicate the order of sites in $H_{\rm ring}$.
We use this convention throughout this Letter.
(b) Antiferromagnetic model on a square lattice. The interaction along the 
diagonal edge is smaller than that along the vertical or horizontal edge. 
(c) A portion of the triangular lattice with the 120$^\circ$ order.}
\label{fig: FM_triangle}
\end{center}
\end{figure}
As an example, we consider the ferromagnetic Heisenberg model on a triangular lattice 
shown in Fig. \ref{fig: FM_triangle}(a) with an ordered moment $S_0$ along $\hat{y}$.
In this case, the quadratic terms in $\delta \vec S$ from 
Eq. (\ref{ring_exchange}) always cancel. 
To explain this, let us focus on the edge $\langle jk \rangle$. From the upper triangle, 
this edge gives $S_0(\delta \vec S_j \times \delta \vec S_k)^y$. On the other hand, 
from the lower triangle, it gives $S_0(\delta \vec S_k \times \delta \vec S_j)^y$ 
which cancels out the former one. Since such a cancellation occurs on any edge, 
$H_{\rm ring}$ in Eq. (\ref{ring_exchange}) does not contribute to the spin-wave Hamiltonian 
to quadratic order. This observation leads us to conclude that such a cancellation 
occurs for any ferromagnetic model where each edge is shared by the 
equivalent cells such as plaquettes and triangles. 
A similar cancellation occurs for 
certain antiferromagnetic systems.
An example is shown in Fig. \ref{fig: FM_triangle}(b). 
In this example, there are several different types of ring exchange
processes, but again the cancellation between the cells sharing a link
occurs for the collinear antiferromagnetic configuration.
Finally, for noncollinear spin structures, we 
considered 
the 
120$^\circ$ magnetic order on a triangular lattice (Fig. \ref{fig: FM_triangle}(c))
and conclude that the cancellation 
again 
occurs since 
ordered components of $\vec S_i$ and that of $\vec S_{\bar i}$ are the same as 
shown in Fig. \ref{fig: FM_triangle}(c).


{\it Intrinsic thermal Hall effect in the spin-wave approximation}. ---
Once we understand the principles of the cancellation, it is rather easy to 
find an example where it does not occur, namely, when the link is shared 
by 
inequivalent 
cells.
An example is the ferromagnetic model on the kagom${\rm \acute{e}}$ lattice.
In this case, the spin wave Hamiltonian is influenced by the 
magnetic flux $\Phi$. 
We will develop below a theoretical formalism to calculate the thermal Hall conductivity in terms of 
the Kubo formula. 
For this purpose, we consider a general Hamiltonian for noninteracting 
bosons which can be regarded as a spin-wave Hamiltonian within 
quadratic order 
in $\delta \vec S$:
\begin{equation}
H=\sum_{j,\alpha} h({\vec R_{j\alpha}}), ~  h({\vec R_{j\alpha}})
=\frac{1}{2} \sum_{\vec \delta_\alpha} t_{\vec \delta_\alpha} 
b^\dagger_{\vec R_{j\alpha} +{\vec \delta_\alpha}} b_{\vec R_{j\alpha}}+{\rm h.c.}, \nonumber
\end{equation}
where $b_{\vec R_{j\alpha}}$ annihilates a boson at the lattice point of the 
$\alpha$-th site in the $j$-th unit cell and $\vec \delta_\alpha$ are vectors 
connecting $\vec R_{j\alpha}$ and its neighboring sites. The hopping 
$t_{{\vec \delta_\alpha}}$ is in general 
complex.
In momentum space, the Hamiltonian is written as
$H=\sum_{\vec k} b^\dagger_\alpha(\vec k)
{\cal H}_{\alpha,\beta}(\vec k) b_\beta(\vec k)$, 
where $b_\alpha(\vec k)$ is a Fourier transform of $b_{\vec R_{j\alpha}}$ and repeated indices are summed over. 
Using the local energy operator $h(\vec R_{j \alpha})$, the average energy current 
density is defined by
\begin{equation}
{\vec j}_E \equiv i [H, \sum_{j \alpha} {\vec R}_{j \alpha} h({\vec R}_{j \alpha})]/V,
\end{equation}
where $V$ is the total volume~\cite{Mahan}. Using the Fourier transform of 
$\sum_\alpha h(\vec R_{j \alpha})$ defined by 
$h(\vec q) = \sum_{j,\alpha} e^{i\vec q \cdot \vec R_{j\alpha}} h(\vec R_{j\alpha})$, 
the energy current density is rewritten as 
$\vec j_E = \partial_{\vec q}[h(0), h(\vec q)]|_{\vec q=0}/V$. 
Using this fact, the following convenient expression for $\vec j_E$ is obtained: 
\begin{equation}
{\vec j}_E =\frac{1}{2V} \sum_{\vec k} 
b^\dagger_{\alpha}(\vec k)\left(\partial_{\vec k} 
{\cal H}(\vec k)^2 \right)_{\alpha \beta} b_\beta(\vec k), 
\label{energy_current}
\end{equation}
where the differential operator $\partial_{\vec k}$ acts only on 
${\cal H}(\vec k)^2$. 
We introduce the spin wave basis $|u_\alpha(\vec k)\rangle$ which diagonalizes ${\cal H}(\vec k)$ with 
eigenvalues $\omega_\alpha(\vec k)$.  It is important to note that even in this basis $\vec j_E$ 
is not diagonal.  In addition to the expected diagonal term 
$\frac{d\omega_\alpha}{d\vec k}\omega_\alpha(\vec k)$, 
there are off-diagonal terms which can be thought of as arising from anomalous velocities.  
As we see below, these terms are responsible for $\kappa^{xy}$, just as in the case of 
the intrinsic anomalous Hall effect in metals \cite{Haldane_AHE}.

Starting from the Kubo formula, 
the following expression analogous to 
the TKNN formula~\cite{TKNN} can be obtained 
for the thermal Hall conductivity $\kappa^{xy}$:
\begin{eqnarray}
\kappa^{xy}= &-&\frac{1}{2T} {\rm Im} \sum_{\alpha} 
\int_{\rm BZ}\frac{d^2 k}{(2\pi)^2} n_\alpha(\vec k) \nonumber \\
&\times & \left\langle \partial_{k_x} u_\alpha(\vec k)  
\big| ({\cal H}(\vec k)+\omega_\alpha(\vec k))^2 \big| \partial_{k_y} u_\alpha (\vec k) \right\rangle,
\label{gauge_inv_formula}
\end{eqnarray}
for the non-interacting spin waves (free bosons) where 
the integral is over the 
Brillouin zone (BZ), and $n_\alpha(\vec k)=(e^{\beta \omega_\alpha(\vec k)}-1)^{-1}$ 
is the Bose distribution function. 
We can show that the integrand in Eq. (\ref{gauge_inv_formula}) is independent of
a $\vec k$-dependent phase change of the spin wave basis functions $|u_\alpha (\vec k)\rangle$ 
in a similar way as for the orbital magnetization~\cite{Ceresoli_PRB}. 

\begin{figure}[tb]
\begin{center}
\vspace{.5cm}
\hspace{-.0cm}\includegraphics[width=\columnwidth]{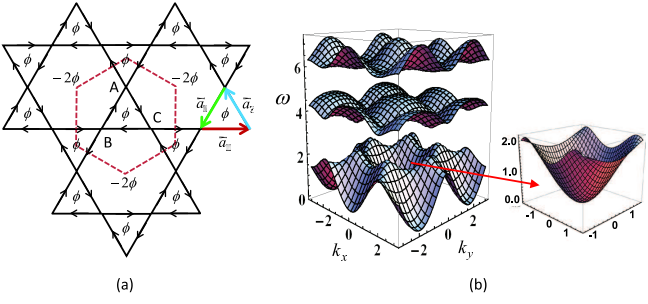}
\vspace{.0cm}
\caption{(a) The unit cell of the kagom${\rm \acute{e}}$ lattice 
(the area enclosed by the dotted line). 
The vectors $\vec a_i$ ($i=1,2$, and $3$) correspond to green, red, 
and blue arrows, respectively. 
The arrows on the edges indicate the sign of the phase factor $e^{i\phi/3}$ 
(see main text). The fictitious fluxes through triangle and hexagon are $\phi$ and $-2\phi$, respectively.
(b) Dispersions of bands in the case of $\phi=\pi/3$ with $JS=1$. 
The right figure shows the enlargement of dispersion around $\omega=0$ where the lowest band has a quadratic dispersion, i.e., $\omega_1(\vec k)\sim JS(k^2_x +k^2_y)$.}
\label{fig: kagome_BZ}
\end{center}
\end{figure}

{\it Thermal Hall effect in kagom${\rm \acute{e}}$ ferromagnet}. ---
We now apply the above formula to the ferromagnetic model on the kagom${\rm \acute{e}}$ 
lattice. The Hamiltonian is given by
$H=\sum_{\bigtriangleup, \bigtriangledown} h_{\bigtriangleup}
+h_{\bigtriangledown}$ with
\begin{equation}
h_{\bigtriangleup/\bigtriangledown}=-J(\vec S_i\cdot \vec S_j 
+ \vec S_j \cdot \vec S_k +\vec S_k \cdot \vec S_i) 
- \frac{K}{S} \vec S_i \cdot (\vec S_j \times \vec S_k), \nonumber
\end{equation}
where $-J$ is the ferromagnetic exchange coupling,
$K$ is proportional to $\sin\Phi$ according to Eq. (\ref{ring_exchange}), 
and the sum is taken over 
all the triangles in the kagom${\rm \acute{e}}$ lattice (see Fig. \ref{fig: kagome_BZ}(a)). 
Again, note that the sites $i$, $j$, and $k$ form a triangle ($\bigtriangleup$ or 
$\bigtriangledown$) in a counterclockwise way. 
Using the Holstein-Primakoff transformation [$S^+_j=(2S-n_j)^{1/2}b_j, S^-_j
=b^\dagger_j (2S-n_j)^{1/2}, S^z_j=S-n_j$ with $n_j=b^\dagger_j b_j$], we obtain
the spin-wave Hamiltonian as
\begin{equation}
H_{\rm SW}=4JS\sum_j n_j -S \sqrt{J^2+K^2}\sum_{\langle j,k\rangle} (e^{-i\phi/3}
b^\dagger_j b_k+{\rm h.c.}),\nonumber
\end{equation}
where $\tan (\phi/3)=K/J$ and the sum is taken over all the nearest 
neighbor bonds. 
This Hamiltonian contains a non-trivial phase factor and 
is quite similar to the electronic model considered by Ohgushi {\em et al}.~\cite{Ohgushi-Murakami-Nagaosa}. 
The Fourier transform of the Hamiltonian is given by 
${\cal H}(\vec k)=4JS-2JS(\cos(\phi/3))^{-1}\Lambda(\vec k, \phi)$ with
\begin{eqnarray}
\Lambda(\vec k, \phi)&=&
\left(\begin{array}{ccc}
0 & \cos k_1 e^{-i\phi/3} & \cos k_3 e^{i\phi/3} \\
\cos k_1 e^{i\phi/3} & 0 & \cos k_2 e^{-i\phi/3} \\
\cos k_3 e^{-i\phi/3} & \cos k_2 e^{i\phi/3} & 0
\end{array}\right), \nonumber
\end{eqnarray}
where $k_i=\vec k \cdot \vec a_i$ with $\vec a_1=(-1/2, -\sqrt{3}/2)$, 
$\vec a_2=(1,0)$, and $\vec a_3=(-1/2, \sqrt{3}/2)$ 
as shown in Fig. \ref{fig: kagome_BZ}(a). 
If the magnetic field is absent, i.e.,  $\alpha=\phi=0$, the upper band is  dispersionless and the other two bands are particle-hole symmetric around $\omega=3JS$. 
The dispersions of three bands ($0 \le \omega_1(\vec k) \le \omega_2(\vec k) 
\le \omega_3(\vec k)$) for $\phi=\pi/3$ are shown in Fig. \ref{fig: kagome_BZ}(b). 
In the limit of low temperature and weak magnetic field, 
the dominant contribution to the integral in Eq. (\ref{gauge_inv_formula}) 
comes from $\alpha=1$ (lowest band) and small $|\vec k|$ due to the Bose 
factor $n_\alpha({\vec k})$. 
By an explicit calculation~\cite{first order phi}, we find that 
$\langle \partial_{k_x}u_1(\vec k) |\partial_{k_y}u_1(\vec k) 
\rangle \sim -i\phi|\vec k|^2/(27\sqrt{3})$ 
around $\vec k=0$ and obtain
\begin{equation}
\kappa^{xy} \sim -\frac{(6JS)^2}{2T} \int^\infty_0 \frac{dk}{2\pi} 
\frac{k}{e^{\beta JSk^2}-1} \left(\frac{-\phi k^2}{27\sqrt{3}}\right) 
=\frac{\pi \phi}{36\sqrt{3}}T, 
\end{equation}
where we have replaced the integration over the BZ with that over all $\vec k$. In this way, 
a non-zero thermal Hall conductivity is indeed realized by the coupling between the scalar 
chirality and the magnetic field in the 
ferromagnetic 
kagom${\rm \acute{e}}$ lattice. 
The obtained linear dependence of $\kappa^{xy}$ on $T$  
is due to the dimensionality ($D=2$) and the quadratic dispersion around $\omega(\vec k)=0$. 


{\it Thermal Hall effect in quantum spin liquids}. ---
As discussed above, the spin Hamiltonian contains the magnetic 
flux $\Phi$, and the spin waves or magnons 
are influenced by the magnetic field only through the off-diagonal 
matrix elements of the thermal currents, corresponding to the 
intrinsic anomalous Hall effect. This is in sharp contrast to the case
of electrons which is coupled to the vector potential ${\vec A}$, and
the usual Hall effect due to the Lorentz force occurs there. 
In this respect, it is interesting to note that the 
spin operator ${\vec S}_i$ can be represented by the 
fermion operators $f^\dagger_{i \sigma}$ and $f_{i \sigma}$ (called {\it spinons} with spin 1/2) as 
${\vec S}_i = \sum_{\sigma,\sigma'} f^\dagger_{i \sigma} {\vec \sigma} 
f_{i \sigma'}/2$ (${\vec \sigma}=(\sigma_x, \sigma_y, \sigma_z)$: Pauli matrices,
$\sigma, \sigma'=\uparrow, \downarrow$) with the constraint 
$\sum_{\sigma} f^\dagger_{i \sigma} f_{i \sigma} =1$.
In this representation, the exchange interaction 
$J {\vec S}_i \cdot {\vec S}_j$ can be written as
$- J \chi^\dagger_{ij} \chi_{ij}/2$ with 
$\chi_{ij} = \sum_\sigma f^\dagger_{i \sigma} f_{j \sigma}$.
This $\chi_{ij}$ is called the order parameter of 
the resonating valence bond (RVB) which describes the 
singlet formation between the two spins ${\vec S}_i$ and
${\vec S}_j$. In the mean field approximation for 
$\chi_{ij}$, the free fermion model for $f^\dagger_{i \sigma}$, $f_{i \sigma}$
emerges
~\cite{LNW}. 
The phase $a_{ij}$ of the order parameter $\langle \chi_{ij} \rangle = |\langle \chi_{ij} \rangle|e^{i a_{ij}}$
and the Lagrange multiplier $a^0$ to impose the constraint
above constitutes the gauge field, which is coupled to the spinons. 
In the confining phase of this gauge field, two spinons are bound to form a magnon. 
On the other hand, in the deconfining phase, the spinons behave as nearly free quasi-particles.
The latter case is realized in some of the quantum 
spin liquid states~\cite{LNW}. 
Similar state has been obtained also for the
Hubbard model~\cite{SSLee_PALee_PRL}, which contains
the gapped charge excitations. 
This charge degrees of freedom is represented by the U(1) phase factor $e^{i \theta_i}$, i.e., the electron operator $c_{i \sigma}$ is decomposed into the product $f_{i \sigma} e^{- i \theta_i}$, which is coupled to $a_{ij}- A_{ij}$ where $A_{ij}$ is the vector potential (Peierls phase) corresponding to the magnetic flux $\Phi$ \cite{SSLee_PALee_PRL}. 
Then, the Maxwell term
${\cal L}_g = (1/g) \int dr \sum_{\mu \nu} ({\cal F}_{\mu \nu} - F_{\mu \nu})^2$ 
($g$: coupling constant, 
${\cal F}_{\mu \nu} = \partial_\mu a_\nu - \partial_\nu a_\mu$,
$F_{\mu \nu} = \partial_\mu A_\nu - \partial_\nu A_\mu$,
) is generated by integrating over the charge degrees of freedom.
To summarize, the spinons are described by the Lagrangian
\begin{eqnarray}
{\cal L}&=& \sum_{j,\sigma} f^\dagger_{j\sigma} (\partial_\tau-ia^0_j -\mu) f_{j \sigma} 
\nonumber \\ 
&-&\sum_{j,k} t_f e^{ia_{jk} } f^\dagger_{j\sigma} f_{k\sigma}
+ {\cal L}_g,   
\end{eqnarray}
Following the previous works~\cite{Motrunich1, SSLee_PALee_PRL}, 
we take the spinon metal with a Fermi surface as a candidate for the 2D quantum spin liquid
realized in $\kappa$-(ET)$_2$Cu$_2$(CN)$_3$\cite{M_Yamashita}.
In a magnetic field $F_{xy} = B_z$, the average of the 
gauge flux $\langle {\cal F}_{xy} \rangle = c F_{xy}$ is induced
with $c$ a constant of the order of unity  
because of the coupling between ${\cal F}_{xy}$ and $F_{xy}$ in 
${\cal L}_g$~\cite{Motrunich1, LNW}. 
Therefore, the spinons are subject to the {\it effective
magnetic field} $\langle {\cal F}_{xy} \rangle$ and to the Lorentz force.  
 
Let us first estimate the spinon lifetime $\tau$ from the recent thermal 
transport measurements in that material. 
The longitudinal thermal conductivity is obtained from the Wiedemann-Franz law by assuming a Fermi liquid of spinons: 
\begin{equation}
\kappa^{xx}_{\rm sp}=2 \frac{\pi^2}{3} \left(\frac{\varepsilon_{\rm F}}{\hbar}\tau \right) 
\frac{k^2_{\rm B}T}{h} \cdot \frac{1}{d},
\end{equation}
where $\varepsilon_{\rm F}$ is the Fermi energy and $d\sim 16\AA$ is the interlayer distance. 
After a subtraction of the phonon contribution, 
$\kappa^{xx}_{\rm sp}$
is estimated to be
$\sim$0.02 WK$^{-1}$m$^{-1}$ at $T=0.3$ K~\cite{M_Yamashita}.  
We obtain $\varepsilon_{\rm F} \tau /\hbar = 56.5$ and with 
$\varepsilon_{\rm F}=J\sim250$ K, 
we 
estimate 
$\tau \sim 1.72 \times 10^{-12}$ s. 
Next we examine $\kappa^{xy}_{\rm sp}$. As has been shown in \cite{Motrunich1}, the gauge 
flux for spinons is comparable to the applied magnetic 
flux and hence $\kappa^{xy}_{\rm sp} \sim (\omega_c \tau)\kappa^{xx}_{\rm sp}$, where 
$\omega_c=eB/m_c$ is the cyclotron frequency with the effective mass of spinon $m_c$. 
Estimating $m_c \sim \hbar^2/(Ja^2)$ with assuming the lattice spacing $a \sim 10 \AA$, 
we obtain $\omega_c \tau \sim 0.086\times B$ with $B$ being measured in Tesla. 
Therefore, the thermal Hall angle $\kappa^{xy}_{\rm sp}/\kappa^{xx}_{\rm sp} \sim \omega_c \tau$ 
becomes of the order of 0.1, which is easily measurable, with a weak magnetic field $B \sim 1$T 
such that the spin-liquid ground state is not disturbed. 
Also note that compared with the intrinsic thermal Hall effect discussed 
above, the magnitude of this Lorentz-force driven thermal Hall conductivity is
much larger by the factor of $\sim (\varepsilon_{\rm F} \tau/\hbar)^2$. 
Therefore, the 
observation of 
the thermal Hall effect 
is 
a clear signature of such deconfined
spinons in the spin liquid, and experiments on  $\kappa$-(ET)$_2$Cu$_2$(CN)$_3$
is highly desirable.

Another important difference between the spinon contribution and the intrinsic term is that the spinons are diffusive and see the field $\vec A$.  Thus in a small sample one can expect mesoscopic effects such as universal conductance fluctuations of the thermal conductivity as a function of $B$.  Using the Wiedemann-Franz law, we expect the relative fluctuation in $\kappa^{xx}$ and $\kappa^{xy}$ to be of order $\hbar/(\varepsilon_{\rm F}\tau)$ for each coherent volume with dimension $\sqrt{\ell \ell_{in}}$ where $\ell = v_{\rm F}\tau$ and $\ell_{in} = v_{\rm F}\tau_{in}$ and $\tau_{in}$ is some inelastic scattering time much longer than $\tau$ at low temperatures.  The fluctuation is reduced by $\sqrt N$ if the sample contains $N$ coherent volumes.  We estimate the elastic mean free path  $\ell$ to be 400 $\rm{\AA}$, so that at low temperatures this effect may be observable in micron-scale samples.	

In conclusion, we have studied theoretically the thermal Hall effect
in the quantum spin systems induced by the external magnetic field.
There are three cases, i.e., (i) no thermal Hall effect,
(ii) intrinsic thermal Hall effect by the magnons, and (iii) large thermal Hall effect
due to the Lorentz force. (i) corresponds to the most of the conventional
(anti)ferromagnets on triangular and square and cubic lattices, while
(ii) to the magnets on a particular lattice structure such as kagom${\rm \acute{e}}$, and
(iii) to the spin liquid with deconfined spinons. Therefore, the thermal Hall 
effect offers a unique experimental method to gain an important insight
on the ground state/low energy excitations of the quantum magnets.
 
The authors are grateful to N.~P.~Ong, T.~Senthil, Y.~Taguchi, Y.~Tokura, and S.~Yamashita 
for their valuable comments and discussions. 
This work is partly supported in part 
by Grant-in-Aids (No. 17105002, No. 19048015, No. 19048008)  
from the Ministry of Education,
Culture, Sports, Science and Technology of Japan.  
PAL acknowledges support by NSF DMR-0804040.

\begin{widetext}
\vspace*{0.3cm}
\begin{center}
\Large 
Supplementary Items for {\it Theory of Thermal Hall Effect in Quantum Magnets}
\end{center}
\subsection{Off-diagonal terms in energy current density}
In this section, we give an explicit expression for the energy current density $\vec j_E$ in the basis where the Hamiltonian is diagonal. As we have explained in the main text, $\vec j_E$ is not diagonal in this basis. 
We start with the non-interacting Hamiltonian in the Fourier space:
\begin{equation}
H=\sum_{\vec k, \alpha,\beta}b^\dagger_\alpha(\vec k){\cal H}_{\alpha,\beta}(\vec k) b_\beta(\vec k),
\end{equation}
where $b_\al(\vec k)$ denotes the Fourier transform of $b_{\vec R_{j\al}}$ and repeated indices are summed over.
The single-particle Hamiltonian ${\cal H}(\vec k)$ can be diagonalized by the unitary matrix $g(\vec k)$ as $g^\dagger(\vec k) {\cal H}(\vec k) g(\vec k)=\Omega(\vec k)$, where $\Omega(\vec k)={\rm diag}(\omega_1(\vec k), ..., \omega_n(\vec k))$.  Using $g(\vec k)$ we define a new basis of bosons as
\begin{equation}
\gamma_i (\vec k) =\sum_j g\dag(\vec k)_{ij} b_j(\vec k).
\end{equation}
In this basis, $H$ is written as $H=\sum_{\vec k} \omega_\alpha(\vec k) \gamma^\dagger_\alpha(\vec k) \gamma_\alpha (\vec k)$. 

Let us now see the off-diagonal structure of $\vec j_E$ in the basis of $\gamma_\al(\vec k)$. 
This is clarified by introducing the Berry connection (or Maurer-Cartan 1-form) defined by $\vec \Theta(\vec k)=g\dag(\vec k) \pa_{\vec k} g(\vec k)$. Using $\vec \Theta$, the energy current density (Eq. (4) in the main text) is written as
\begin{equation}
\vec j_E = \frac{1}{2V} \sum_{\vec k} \gamma\dag_\alpha (\vec k) \left( \frac{\partial}{\partial \vec k}\Omega(\vec k)^2+[\vec \Theta(\vec k), \Omega(\vec k)^2] \right)_{\alpha \beta} \gamma_\beta(\vec k).
\end{equation}
Here we have used the relation $(\partial_{\vec k} g\dag (\vec k)) g(\vec k)=- g\dag(\vec k) (\partial_{\vec k} g(\vec k))$. 
The first term in the bracket is a diagonal one corresponding to $\frac{\pa\omega_\al (\vec k)}{\pa {\vec k}} \omega_\al(\vec k)$. On the other hand, the second term represented by the commutation relation gives the off-diagonal elements of the energy current density. Similarly to the case of the intrinsic anomalous Hall effect, this off-diagonal term can be regarded as arising from anomalous velocities. 
The thermal conductivity tensor can be obtained by substituting the above expression for $\vec j_E$ into the Kubo formula:
\begin{equation}
\kappa^{xy}=\frac{V}{T}\int^{\infty}_0 dt \int^{\beta}_0 d\lambda
\langle j^x_E(-i\lambda) j^y_E(t) \rangle_{\rm th},
\end{equation}
where $\beta=1/T$ is the inverse temperature and  $\langle ... \rangle_{\rm th}$ denotes thermal average. Along the same lines as the derivation of the TKNN formula~\cite{TKNN}, one can obtain Eq. (\ref{gauge_inv_formula}).

\subsection{A detailed derivation of the thermal Hall conductivity in the Kagom${\acute {\rm e}}$ ferromagnet}
\subsubsection{Eigenvalue problem}
\hspace{5mm}In the main text, we have studied the thermal Hall effect in the kagom${\acute {\rm e}}$ ferromagnet. To obtain the spin-wave spectrum and eigen-modes, we need to solve the eigenvalue problem associated with the following matrix:
\begin{equation}
\Lambda(\vec k, \phi)=
\left(\begin{array}{ccc}
0 & \cos(\vec k \cdot \vec a_1)e^{-i\phi/3} & \cos(\vec k \cdot \vec a_3)e^{i\phi/3} \\
\cos(\vec k \cdot \vec a_1)e^{i\phi/3} & 0 & \cos(\vec k \cdot \vec a_2)e^{-i\phi/3} \\
\cos(\vec k \cdot \vec a_3)e^{-i\phi/3} & \cos(\vec k \cdot \vec a_2)e^{i\phi/3} & 0
\end{array}\right).
\end{equation}
Before study this matrix, we consider a more general matrix
\begin{equation}
M=\left(\begin{array}{ccc}
0 & a & c^* \\
a^* & 0 & b \\
c & b^* & 0
\end{array}\right)
\end{equation}
and its secular equation given by
\begin{equation}
\lambda^3=(|a|^2+|b|^2+|c|^2)\lambda+(abc+a^*b^*c^*).
\label{sec}
\end{equation}
Now we apply Viete's solution of cubic equation. To do so, we compare
Eq. (\ref{sec}) with the following identity:
\begin{equation}
\cos^3 \left(\frac{\theta}{3}\right)=\frac{3}{4}\cos\left(\frac{\theta}{3}\right)+\frac{1}{4}\cos\theta.
\end{equation} 
Then, we obtain one of the solutions $\lambda_1=2A\cos(\theta/3)$ with $\cos \theta=B/(2A)$ ($0 \le \theta \le \pi$), where $A$ and $B$ are defined through the following relations:
\begin{equation}
3A^2=|a|^2+|b|^2+|c|^2,~~~~~A^2 B=abc+a^*b^*c^*.
\end{equation}
The other two solutions can also be written in terms $\theta$ as
$\lambda_{2}$ or $\lambda_3=2 A\cos[(\theta \pm 2\pi)/3]$.
The normalized eigenvector corresponding to $\lambda_n$ ($n=1,2,3$) is explicitly given by $W^{-1}_n (\lambda^2_n-|b|^2, \lambda_n a^*+bc, \lambda_n c+a^* b^*)^T$ where $W^2_n=3[2A^2 \lambda^2_n+(A^2-\lambda^2_n)|b|^2+\lambda_n A^2 B]$. 

Let us now apply the above general result to our specific problem of $\Lambda(\vec k, \phi)$. We identify $3A^2$ with $1+f(\vec k)$, and
$A^2 B$ with $f(\vec k)\cos(\phi)$, where $f(\vec k)=2\prod^3_{j=1}\cos(\vec k \cdot \vec a_j)$. 
Then, the eigenvalues of $\Lambda(\vec k, \phi)$ are given by
\begin{equation}
\lambda_n(\vec k, \phi)=2 \sqrt{\frac{1+f(\vec k)}{3}} \cos\left(\frac{\theta_n(\vec k)}{3}\right),~~~~~(n=1,2,3)
\label{lam_n}
\end{equation}
with
\begin{equation}
\cos \theta_1(\vec k)=\sqrt{\frac{27}{4}\frac{(f(\vec k)\cos\phi)^2}{(1+f(\vec k))^3}},
~~~~~\theta_{2}(\vec k)=\theta_1(\vec k)+ 2\pi,~~~~~\theta_3(\vec k)=\theta_1(\vec k)-2\pi.
\label{theta_2_3}
\end{equation}
It is useful to note that $1+f(\vec k)=\sum^3_{j=1} \cos^2(\vec k \cdot \vec a_j)$. The normalized eigenvector corresponding to $\lambda_n(\vec k, \phi)$ is written as
\begin{equation}
|u_n(\vec k)\ra = \frac{1}{W_n(\vec k, \phi)}\left(\begin{array}{c}
\lambda_n(\vec k, \phi)^2-\cos^2 k_2 \\
e^{i\phi/3} \lambda_n(\vec k, \phi)\cos k_1 +e^{-2i\phi/3}\cos k_2 \cos k_3 \\
e^{-i\phi/3} \lambda_n(\vec k, \phi)\cos k_3 +e^{2i\phi/3}\cos k_1 \cos k_2 
\end{array}\right),
\end{equation}
where $W_n(\vec k, \phi)^2=2(1+f(\vec k))\lambda_n(\vec k)^2 + (1+f(\vec k)-3\lambda_n(\vec k)^2)\cos^2 k_2 + 3\lambda_n(\vec k)f(\vec k)\cos \phi$.
As for the ferromagnetic Kagom${\acute{\rm e}}$ model, the Hamiltonian
is defined by ${\cal H}(\vec k)=4JS-2JS(\cos(\phi/3))^{-1}\Lambda(\vec k, \phi)$ and hence the eigen-energies are given by
\begin{equation}
\omega_n(\vec k) = 4JS \left( 1-\sqrt{\frac{1+f(\vec k)}{3}} \frac{\cos(\theta_n(\vec k)/3)}{\cos(\phi/3)} \right).
\end{equation}

\subsection{Derivation of Eq. (6)}
\hspace{5mm}
We now derive Eq. (6) in the main text from the general formula for $\kappa^{xy}$ (see Eq. (5)). 
In the limit of low temperature and weak magnetic field, the contribution from the lowest band ($\alpha=1$) dominates in RHS of Eq. (5) due to the Bose factor $n_\alpha(\vec k)$. Therefore, we approximate Eq. (5) as
\begin{equation}
\kappa^{xy} \sim -\frac{1}{2T} {\rm Im}  
\int_{\rm BZ}\frac{d^2 k}{(2\pi)^2} n_1 (\vec k) 
\left\langle \partial_{k_x} u_1 ({\vec k})  
\big| ({\cal H}({\vec k})+\omega_1 ({\vec k}))^2 \big| \partial_{k_y} u_1 ({\vec k}) \right\rangle. 
\end{equation}
Inserting the resolution of the identity at each $\vec k$, i.e., $1=\sum^3_{\beta=1}|u_\beta(\vec k)\rangle \langle u_\beta (\vec k)| $, we can rewrite the above expression as
\begin{equation}
\kappa^{xy} \sim -\frac{1}{2T} \sum_{\beta=2,3}  
\int_{\rm BZ}\frac{d^2 k}{(2\pi)^2} n_1 ({\vec k}) (\omega_1({\vec k})+\omega_\beta({\vec k}))^2~
{\rm Im}\left[
\left\langle \partial_{k_x} u_1 ({\vec k})  
\Big| u_\beta({\vec k}) \right\rangle \left\langle u_\beta(\vec k) \Big| \partial_{k_y} u_1 (\vec k) \right\rangle
\right].
\label{eq_6_corres}
\end{equation}
Note that we have omitted $\beta=1$ from the summation above since it can be shown that $\langle \partial_{k_x}u_1 (\vec k)| u_1 (\vec k)\rangle \langle u_1(\vec k)| \partial_{k_y} u_1(\vec k)\rangle$ is real using the fact $|u_1(\vec k) \rangle$ is normalized, i.e., $\langle \partial_{\vec k} u_1(\vec k)| u_1 (\vec k)\rangle +\langle u_1(\vec k)| \partial_{\vec k} u_1(\vec k)\rangle =0$. 
In the low temperature limit, the dominant contribution to the integral in Eq. (\ref{eq_6_corres}) comes from small $|\vec k|$ due to the Bose factor. 
In the vicinity of $\vec k=\vec 0$, one can expand $\cos \theta_1(\vec k)$ in Eq. (\ref{theta_2_3}) and $\lambda_1(\vec k, \phi)$ in Eq. (\ref{lam_n}) with respect to $\vec k$ as
\begin{equation}
\cos \theta_1(\vec k) \sim \cos\phi (1+ O(k^4)),~~
\lambda_1(\vec k, \phi) \sim 2 \cos (\phi/3) \left(1-\frac{1}{6}\sum^3_{j=1}k^2_j+O(k^4) \right).
\label{lambda1}
\end{equation}
Therefore, in the vicinity of $\vec k = \vec 0$, $\omega_1(\vec k)\sim JS|\vec k|^2$. 
Furthermore, using Eq. (\ref{theta_2_3}) and (\ref{lambda1}), we find $\omega_1(\vec k)+\omega_2(\vec k) \sim 2JS(3-\sqrt{3} \tan(\phi/3)+O(k^2))$
and
$\omega_1(\vec k)+\omega_3(\vec k)\sim 2JS(3+\sqrt{3} \tan(\phi/3)+O(k^2))$. 
Then, we approximate $\omega_1+\omega_2$ and $\omega_1+\omega_3$ by $6JS$ and rewrite Eq. (\ref{eq_6_corres}) as 
\begin{eqnarray}
\kappa^{xy} &\sim & -\frac{1}{2T}   
\int_{\rm BZ}\frac{d^2 k}{(2\pi)^2} n_1 ({\vec k}) (6JS)^2~
\sum_{\beta=2,3}{\rm Im}\left[
\left\langle \partial_{k_x} u_1 (\vec k)  
\Big| u_\beta(\vec k) \right\rangle \left\langle u_\beta(\vec k) \Big| \partial_{k_y} u_1 (\vec k) \right\rangle
\right] \nonumber \\
&=& -\frac{(6JS)^2}{2T} \int_{\rm BZ} \frac{d^2 k}{(2\pi)^2} n_1 ({\vec k})
\sum^3_{\beta=1}{\rm Im}\left[
\left\langle \partial_{k_x} u_1 (\vec k)  
\Big| u_\beta(\vec k) \right\rangle \left\langle u_\beta(\vec k) \Big| \partial_{k_y} u_1 (\vec k) \right\rangle
\right] \nonumber \\
&=& -\frac{(6JS)^2}{2T} \int_{\rm BZ} \frac{d^2 k}{(2\pi)^2} n_1 ({\vec k})~
{\rm Im} \left[ \left\langle \partial_{k_x} u_1 (\vec k) \Big|
\partial_{k_y} u_1 (\vec k) \right\rangle \right].
\label{eq_6_corres2}
\end{eqnarray}
Here, we have again used the fact that $\langle \partial_{k_x}u_1 (\vec k)| u_1 (\vec k)\rangle \langle u_1(\vec k)| \partial_{k_y} u_1(\vec k)\rangle$ is real. Note that if we take into account the $\phi$- and $k$-dependencies of $\omega_1+\omega_i$ ($i=2, 3$), they just give higher order terms in $\phi$ and $T$. (As we will see in the next paragraph, ${\rm Im} [ \langle \partial_{k_x} u_1 (\vec k) |\partial_{k_y} u_1 (\vec k) \rangle ]$ is already first order in $\phi$. )

For our purpose to obtain the thermal Hall conductivity, we need to calculate the imaginary part of $\la \pa_{k_x}u_1(\vec k)|\pa_{k_y}u_1(\vec k) \ra$. Now we consider the perturbative expansion of $|u_1(\vec k)\ra$ with respect to $\phi$ and keep up to the first order terms in $\phi$. The zeroth and the first order wavefunctions are given by
\begin{equation}
|u^{(0)}_1(\vec k)\ra = \frac{1}{W_1(\vec k,0)} \left(\begin{array}{c}
\lambda_1(\vec k,0)^2-\cos^2 k_2 \\ \lambda_1(\vec k,0)\cos k_1+\cos k_2 \cos k_3 \\ \lambda_1(\vec k,0)\cos k_3 + \cos k_1 \cos k_2 \end{array}\right)
\end{equation}
and
\begin{equation}
|u^{(1)}_1(\vec k)\ra = \frac{i(\phi/3)}{W_1(\vec k,0)}\left(\begin{array}{c}
0 \\ \lambda_1(\vec k,0)\cos k_1 -2\cos k_2 \cos k_3 \\ -\lambda_1(\vec k,0)\cos k_3 + 2\cos k_1 \cos k_2
\end{array}\right),
\end{equation}
respectively. Then, from the Taylor expansion around $\vec k=\vec 0$, we obtain
\begin{eqnarray}
{\rm Im}[\la \pa_{k_x}u_1(\vec k)|\pa_{k_y}u_1(\vec k) \ra]
&\sim &{\rm Im}[ \la \partial_{k_x} u^{(0)}_1(\vec k)|\partial_{k_y} u^{(1)}_1(\vec k) \ra + \la \partial_{k_x} u^{(1)}_1(\vec k)|\partial_{k_y} u^{(0)}_1(\vec k) \ra] \nonumber \\
&=&-\frac{\phi}{27\sqrt{3}}(k^2_x+k^2_y).
\label{quad_BC}
\end{eqnarray} 
Therefore, for the first order term in $\phi$, 
the Berry curvature around the point $\vec k=\vec 0$ is quadratic in $|\vec k|$. Substituting Eq. (\ref{quad_BC}) into Eq. (\ref{eq_6_corres2}), we obtain 
Eq. (6) shown in the main text. 
\end{widetext}

\begin{thebibliography}{99}
\bibitem{extrinsic}
Also the extrinsic contribution to the thermal Hall conductivity from the skew scattering
~\cite{Sinitsyn_JPCond}
is expected. However, at the scattering events with the low energy limit,
the $s$-wave scattering is dominant and the skew scattering effect
is expected to be very small.
\end{thebibliography}
\end{document}